      [1999/12/01 v1.4c Il Nuovo Cimento]
\documentclass{cimento}

\usepackage{epsfig}
\usepackage{psfig}

\newcommand{\tev}{\ensuremath{\mathrm{Te\kern -0.1em V}}}
\newcommand{\gev}{\ensuremath{\mathrm{Ge\kern -0.1em V}}}	
\newcommand{\mev}{\ensuremath{\mathrm{Me\kern -0.1em V}}}	
\newcommand{\kev}{\ensuremath{\mathrm{ke\kern -0.1em V}}}	
\newcommand{\massgev}{\mbox{\gev/$c^2$}}			
\newcommand{\massmev}{\mbox{\mev/$c^2$}}			

\newcommand{\cita}[1]{\cite{#1}}

\newcommand{\Yquattros}{\mbox{$\Upsilon$(4S)}}
\newcommand{\Do}{D\O}

\def\babar{\mbox{\slshape B\kern-0.1em{\smaller A}\kern-0.1em B\kern-0.1em{\smaller A\kern-0.2em R}}}
  
\newcommand{\belle}{Belle}

\newcommand{\Bd}{\ensuremath{B^{0}}}
\newcommand{\aBd}{\ensuremath{\overline{B}^{0}}}

\newcommand{\Bs}{\ensuremath{B_{s}^{0}}}
\newcommand{\aBs}{\ensuremath{\overline{B}_{s}^{0}}}
\newcommand{\Bn}{\ensuremath{B^{0}_{(s)}}}
\newcommand{\Lb}{\ensuremath{\Lambda_{b}^{0}}}
\newcommand{\Jpsi}{\ensuremath{J/\psi}}

\newcommand{\Bhh}{\ensuremath{\Bn \rightarrow h^{+}h^{'-}}}

\newcommand{\Bdpipi}{\ensuremath{\Bd \rightarrow \pi^+ \pi^-}}

\newcommand{\BdKpi}{\ensuremath{\Bd \rightarrow K^{+} \pi^-}}

\newcommand{\aBdKpi}{\ensuremath{\aBd \rightarrow K^- \pi^+}}

\newcommand{\BsKpi}{\ensuremath{\Bs \rightarrow K^- \pi^+}}

\newcommand{\aBsKpi}{\ensuremath{\aBs\rightarrow K^+ \pi^-}}

\newcommand{\BsKK}{\ensuremath{\Bs \rightarrow  K^+ K^-}}

\newcommand{\Bspipi}{\ensuremath{\Bs \rightarrow  \pi^+ \pi^-}}

\newcommand{\BdKK}{\ensuremath{\Bd \rightarrow  K^+ K^-}}

\newcommand{\Lbppi}{\ensuremath{\Lambda_{b}^{0} \rightarrow p\pi^{-}}}

\newcommand{\LbpK}{\ensuremath{\Lambda_{b}^{0} \rightarrow pK^{-}}}

\newcommand{\Bmumu}{\ensuremath{\Bn \rightarrow \mu^{+}\mu^{-}}}
\newcommand{\Bdmumu}{\ensuremath{\Bd \rightarrow \mu^{+}\mu^{-}}}
\newcommand{\Bsmumu}{\ensuremath{\Bs \rightarrow \mu^{+}\mu^{-}}}

\newcommand{\BsJpsiPhi}{\ensuremath{\Bs \to  \Jpsi \phi}}

\newcommand{\stat}{\ensuremath{\mathit{~(stat.)}}}		
\newcommand{\syst}{\ensuremath{\mathit{~(syst.)}}}		

\newcommand{\lumifb}{\mbox{fb$^{-1}$}}				

\newcommand{\babelle}{\mbox{$B$-Factories}}

\newcommand{\BR}{\ensuremath{\mathcal{B}}}

\newcommand{\CP}{{\rm CP}}  

\newcommand{\acp}{\ensuremath{{\cal A}_{\CP}}}
\newcommand{\acpbdkpi}{\ensuremath{\acp(\BdKpi)}}

\newcommand{\acpbskpi}{\ensuremath{\acp(\BsKpi)}}

\newcommand{\dedx}{\ensuremath{\mathit{dE/dx}}}

\newcommand{\BsDspi}{\ensuremath{\Bs \rightarrow D^{-}_{s} \pi^+}}

\newcommand{\BsDsK}{\ensuremath{\Bs \rightarrow D^{\mp}_{s} K^{\pm}}}

\newcommand{\Bc}{\ensuremath{B_c}}

\newcommand{\Sigmab}{\ensuremath{\Sigma_b^\pm}}

\newcommand{\Xib}{\ensuremath{\Xi_b^-}}
\newcommand{\XibJpsiXi}{\ensuremath{\Xib\rightarrow\Jpsi\Xi^-}}

\newcommand{\BS}{\ensuremath{B^*}}
\newcommand{\BSS}{\ensuremath{B^{**0}}}

\title{$B$ Physics at the Tevatron}
\author{M.~J.~Morello\from{ins:x} (for the CDF and \Do\ Collaborations)}
\instlist{\inst{ins:x} Universit\`a and I.N.F.N, Sezione di Pisa - Ed. C,
         Polo Fibonacci, \\ Largo B. Pontecorvo, 3 - 56127 Pisa, Italy}
\PACSes{\PACSit{13.20.He}{Decays of bottom mesons}
\PACSit{13.25.Ft}{Decays of charmed mesons}
\PACSit{13.30.Eg}{Hadronic decays}}

\begin{document}

\maketitle

\begin{abstract}
The Fermilab Tevatron offers unique opportunities to perform measurements of the heavier 
$b$-hadrons that are not accessible at the \Yquattros\ resonance. In this summary, 
we describe most important heavy flavor results from D\O\ and CDF collaborations and we discuss 
prospects for future measurements, that could reveal New Physics before 
the start-up of the Large Hadron Collider (LHC). 
\end{abstract}


\section{\Bs\ Mixing}

In the $\Bs$ system, the mass eigenstates $B^0_{sL}$ and $B^0_{sH}$
are admixtures of the flavor eigenstates $\Bs$ and $\aBs$.  This
causes oscillations between the $\Bs$ and $\aBs$ states with a
frequency proportional to the mass difference of the mass eigenstates,
$\Delta m_s\equiv m_H - m_L$. In the Standard Model this effect is
explained in terms of second-order weak processes involving virtual
massive particles that provide a transition amplitude between the
$\Bs$ and $\aBs$ states. The magnitude of this mixing amplitude is
proportional to the oscillation frequency, while its phase,
responsible for CP~violation in $\BsJpsiPhi$ decays, is
$-2\beta_s^{\textit{SM}} = -2\arg\left(-{V_{ts}V_{tb}^*\over
V_{cs}V_{cb}^*}\right)$~\cite{Ref:Nierste}, where $V_{ij}$ are the
elements of the CKM quark mixing matrix.  The presence of physics
beyond the Standard Model could contribute additional processes and
modify the magnitude or the phase of the mixing amplitude.  The recent
precise measurement of the oscillation frequency from CDF
$\Delta M_s = 17.77 \pm 0.10 \stat \pm 0.07 \syst~{\rm ps^{-1}}$ ~\cite{Ref:BsMixing}
indicates that contributions of  New Physics to the magnitude, if any,
are extremely small~\cite{Ref:Ligeti}.  Global fits of experimental
data tightly constrain the CP~phase to small values in the
context of the Standard Model, $2\beta_s^{\textit{SM}}\approx
0.04$~\cite{Ref:HFAG}.  However, New Physics may contribute
significantly larger values~\cite{Ref:Ligeti,Ref:Nierste2}.  The
observed CP~phase can be expressed as $2\beta_s =
2\beta_s^{\textit{SM}} - \phi_s^{\textit{NP}}$, where
$\phi_s^{\textit{NP}}$ is due to the additional processes.  The
decay-width difference between the mass eigenstates,
$\Delta\Gamma_s\equiv\Gamma_{L}-\Gamma_{H}$, is also sensitive to the same
New Physics phase.  If $\phi_s^{\textit{NP}}\gg
2\beta_s^{\textit{SM}}$, we expect $\Delta\Gamma_s =
2|\Gamma_{12}|\cos(2\beta_s)$~\cite{Ref:Nierste2}, where
$|\Gamma_{12}|$ is the off-diagonal element of the $\Bs$-$\aBs$ decay
matrix from the Schroedinger equation describing the time evolution of
$\Bs$ mesons~\cite{Ref:Kuhr}.  

Recent studies of $\BsJpsiPhi$ decays
without identification of the initial flavor of the $\Bs$
meson~\cite{Ref:Kuhr,Ref:D0} have provided information on
$\Delta\Gamma_s$ and have some limited sensitivity to the
CP~phase. 

Assuming CP conservation (a good approximation
for the \Bs\ system in the Standard Model)
CDF  collaboration measured the mean lifetime, $\tau_s=2/(\Gamma_{L}+\Gamma_H)= 
1.52 \pm 0.04\stat \pm 0.02\syst$~ps, and the 
decay-width difference, $\Delta\Gamma_s= 0.076^{+0.059}_{-0.063}\stat 
\pm 0.006\syst$~ps$^{-1}$ \cite{Ref:Kuhr}, of the light and heavy mass eigenstates, 
in \BsJpsiPhi\ decays using $1.7$~\lumifb\ of data. 
The measured value of $\Delta\Gamma_s$ agrees with the expected Standard Model value of 0.096~ps$^{-1}$
\cite{Ref:Nierste}. Assuming CP violation, $2\beta_s$ free to vary in the fit,    
CDF collaboration obtained an agreement at 1.2 Gaussian standard deviations with respect to the 
SM expectation ($\Delta\Gamma_s \approx 0.1$ ps$^{-1}$ and $2\beta_s \approx 0$).

With a similar analysis, using a data sample corresponding to an integrated luminosity of 1.1 fb$^{-1}$,
\Do\ collaboration obtained a similar result. They measured 
$\Delta \Gamma_s = 0.17 \pm 0.09\stat\pm 0.02\syst$~ps$^{-1}$ and 
the CP-violating phase $\phi _{s} = -2\beta_{s}=-0.79 \pm 0.56\stat^{+0.14}_{-0.01}\syst$~\cite{Ref:D0}.
Under the hypothesis of  no CP violation ($\phi _{s} \equiv 0$), \Do\ obtained
$\tau_s =1.52\pm 0.08\stat ^{+0.01}_{-0.03}\syst$~ps and
$\Delta \Gamma_s =0.12 ^{+0.08} _{-0.10} \pm 0.02$ ps$^{-1}$~\cite{Ref:D0}.

CDF reported also the first determination of bounds on the
$\textit{CP}$-violation parameter $2\beta_s$ using $\Bs$ decays in
which the flavor of the bottom meson at production is identified 
(flavor tagging). Such information is necessary to separate the time evolution of mesons
produced as $\Bs$ or $\aBs$.  By relating this time development with
the CP~eigenvalue of the final states, which is accessible
through the angular distributions of the $\Jpsi$ and $\phi$ mesons, it is possible
to obtain direct sensitivity to the CP-violating phase.
This CDF result is based on approximately $2000$ $\BsJpsiPhi$ decays
reconstructed in a 1.35~fb$^{-1}$ data sample. 
Assuming the Standard Model predictions of $2\beta_s$ and $\Delta\Gamma_s$, the
probability of a deviation as large as the level of the observed data
is 15\%, corresponding to 1.5 Gaussian standard deviations \cite{Ref:Kuhr_tagged}.
Treating $\Delta\Gamma_s$ as a nuisance parameter and fitting only for $2\beta_s$, CDF measured 
that $2\beta_s\in[0.32,2.82]$ at the 68\% confidence level~\cite{Ref:Kuhr_tagged}.

An alternative way of accessing $\phi_s$ is through the relation 
$A_{SL}^s = \frac{\Delta \Gamma_s} {\Delta M_s} \tan{\phi_s}$
where the semileptonic charge asymmetry is defined as \cite{Ref:PDG}:
$A_{SL}^s = \frac{ N(\bar B_s^0 \to \ell^+ X) - N(B_s^0 \to \ell^- X) }
 { N(\bar B_s^0 \to \ell^+ X) + N(B_s^0 \to \ell^- X) }$.
\Do\ collaboration measured   $A_{SL}^s$ directly by
 using  all events with at least one muon 
that were consistent with the sequential decay
 $B_s^0 \to \mu \nu D_s$ with $D_s \to \phi \pi$  \cite{Ref:2bs_combination}:
$A_{SL}^s = +0.0245 \pm 0.0193~\stat \pm 0.0035~\syst$,
and by using the same-sign dimuon charge asymmetry \cite{Ref:acp_semiletonic}
$A_{SL}^s = -0.0064 \pm 0.0101$ \cite{Ref:2bs_combination}.
Their combination gives the best estimate of the charge asymmetry in semileptonic $B_s^0$ decays:
$A_{SL}^s = 0.0001 \pm 0.0090$.
\Do\ combined the  measurement of 
the width difference between the light  and  
heavy  $B_s^0$ mass eigenstates
and of the CP-violating mixing phase determined from 
 the time-dependent angular
distributions in the \BsJpsiPhi\ decays along with the charge asymmetry in 
semileptonic  decays. 
With the additional constraint from the
world average of the flavor-specific $B_s^0$  lifetime, \Do\ obtained  
$\Delta \Gamma_s$ = 0.13 $\pm$ 0.09  ps$^{-1}$ and   
$\phi _{s} = -0.70 ^{+0.47} _{-0.39}$.

In a very recent paper, UTfit collaboration combined all the available experimental 
information on \Bs\ mixing, including these very recent results from 
CDF and \Do\ collaborations \cite{Ref:Utfit_phi_s}. 
The phase of the \Bs\ mixing amplitude deviates more than $3\sigma$ from the Standard Model prediction. 
While no single measurement has a $3\sigma$ significance yet, all the constraints 
show a remarkable agreement with the combined result. This is a first evidence 
of physics beyond the Standard Model. This result disfavours New Physics models 
with Minimal Flavour Violation with the same significance. 

CDF and \Do\ have analysed only a first part of data ($\approx 1.1-1.35~\lumifb$). They 
have on tape already $\approx 3~\lumifb$ and   
expect to collect a data sample of $\approx 6~\lumifb$ by the year 2010. The prospects 
for the near future are very interesting, CDF and \Do\  will probe soon  
this first evidence of New Physics measuring a more precise CP-violating phase.

\section{$D^0$ Mixing}

In 2007 the first evidence of charm mixing was presented by 
the \babelle\ \cite{Ref:babar_mixing,Ref:belle_mixing}.
CDF also reported recently an evidence of charm mixing in an analysis of 1.5~\lumifb\ of data.
Charm mixing is highly suppressed compared with that of the bottom and 
strange sectors because, since the charm is an up-type quark, top cannot 
participate in the mixing loop. CDF can exploit its large samples of charm mesons 
($3.044\times10^6$ ``right-sign'' $D^{\star+}\to D^0\pi^+ \to [K^-\pi^+]\pi^+$ and 
$12.7\times 10^3$ ``wrong-sign'' $D^{\star+}\to D^0\pi^+ \to [K^+\pi^-]\pi^+$) making an analysis of such 
small mixing rate variable. The wrong-sign decays are the result of either $D^0$
mixing or doubly Cabibbo suppressed decays. The ratio of the wrong-sign to 
right-sign events as a function of $D^0$ decay time probes the charm sector mixing 
parameters $x^{\prime}$ and $y^{\prime}$: 
$R(t)=R_{D}+y^{\prime}\sqrt{R_{D}}t+(x^{\prime 2}+ y^{\prime 2})t^2/4$. 
CDF collaboration measured the time dependence of this ratio 
in a range of proper decay time between 0.75 and
10 mean $D^0$ lifetimes, obtaining 
the mixing parameters to be 
$R_D = (3.04 \pm0.55) \times 10^{-3}$, $y' = (8.5 \pm 7.6) \times 10^{-3}$, and $x'^2 =
(-0.12 \pm 0.35) \times 10^{-3}$ \cite{Ref:cdf_mixing}.
CDF also reported Bayesian probability contours in the $x'^2-y'$ plane finding
that the data are inconsistent with the no-mixing hypothesis with
a probability equivalent to 3.8 Gaussian standard deviations \cite{Ref:cdf_mixing} in agreement
with \babelle\ results. 

The prospects for the future are very promising, CDF will collect
$\approx 6~\lumifb$ by the year 2010 and could observe for the first time the $D^0$ oscillations.

\section{Search for the rare \Bmumu\ decays}

Processes involving flavor-changing neutral currents (FCNCs) provide excellent
signatures with which to search for evidence of New Physics since they have small branching fractions in the
Standard Model (SM), while New Physics contributions can provide a significant enhancement.
The FCNC decays $\Bs (\Bd) \rightarrow \mu^+\mu^-$ occur in the SM only through higher order diagrams and are
further suppressed by the helicity factor, $(m_\mu/m_B)^2$. 
The $\Bd$ decay is also suppressed with respect to the $\Bs$ decay
by the ratio of CKM elements, $\left|V_{td}/V_{ts}\right|^2$.
The SM expectations for these branching fractions are 
$\BR(\Bsmumu) = (3.42\pm0.54)\times10^{-9}$ and $\BR(\Bdmumu) = (1.00\pm0.14)\times10^{-10}$~\cite{smbr}. 

Enhancements to \Bsmumu\ and \Bdmumu\ occur in many New Physics models.
In general, the search for these rare
decays is central to exploring a large class of New Physics models \cite{tanb,rparity,dark,chi2,nmfv}.

Both CDF and \Do\ collaborations have performed a search for \Bsmumu\ and \Bdmumu\ decays in $p\bar{p}$ collisions
at $\sqrt{s} = 1.96$~TeV using 2~\lumifb\ of integrated luminosity.  
The observed number of \Bs\ and \Bd\ candidates is consistent with background expectations in both 
experiments.   The resulting upper limits on the branching fractions are 
$\BR(\Bsmumu) < 5.8\times10^{-8}$ and $\BR(\Bdmumu) < 1.8\times10^{-8}$ at $95\%$ C.L. for CDF 
\cite{Ref:cdf_Bmumu} and $\BR(\Bsmumu) < 9.3\times10^{-8}$ at $95\%$ C.L. for \Do\ \cite{Ref:D0_Bmumu}.

Tevatron experiments observe no evidence for New Physics and set limits that are the most stringent
to date. 
These limits place further constraints on
new-physics models~\cite{tanb,rparity,dark,chi2,nmfv}, and complement direct searches for New Physics.
We expect the analysis sensitivity to continue to improve as we include larger data sets.

\section{Measurements of \Bhh\  decays}

CDF experiment analysed an integrated luminosity of 1~\lumifb\ sample of pairs of oppositely-charged particles
and reconstructed a sample of 14500 \Bhh\ decay modes (where $h= K~{\rm or}~ \pi$) after the off-line confirmation
of trigger requirements. 
In the off-line analysis, the selection cuts was aimed to  minimize the expected uncertainty of the physics
observables to be measured (through several ``pseudo-experiments'').
Two different sets of cuts have been used, respectively optimized
to measure the \CP\ asymmetry \acpbdkpi\ 
and to improve the sensitivity for discovery and limit setting~\cite{gp0308063} of the not yet observed \BsKpi\ mode.
The resolution in invariant mass and in particle identification (\dedx) is not 
sufficient for separating the individual \Bhh\ decay modes on an event-by-event basis,
therefore a maximum likelihood fit was performed. This combines kinematic and particle identification information
to statistically determine both the contribution of each mode,
and the relative contributions to the CP asymmetries.
Significant signals are seen for \Bdpipi, \BdKpi, and \BsKK\ previously observed by CDF~\cite{paper_bhh}.
Three new rare modes were observed for the first time \BsKpi, \Lbppi\ and \LbpK,
with a significance respectively of $8.2 \sigma$, $6.0 \sigma$ and $11.5 \sigma$. 
No evidence was obtained for \Bspipi or  \BdKK\ mode.

The branching fraction of the newly observed mode 
$\BR(\BsKpi)=(5.0 \pm 0.75 \pm 1.0)\times 10^{-6}$~\cite{morello_beauty} is in agreement with the latest
theoretical expectation \cita{zupan}, which is lower than  the previous predictions \cite{B-N,Yu-Li-Cai}.
For the first time CDF measured in the \Bs\ meson system
the direct \CP\ asymmetry $\acpbskpi=0.39 \pm 0.15 \pm 0.08$~\cite{morello_beauty}.
This value favors a large \CP\ violation in  \Bs\ meson decays, 
although it is also compatible with zero. 
In Refs.~\cite{Gronau-BsKpi,Lipkin-BsKpi}  a robust test of the Standard Model or a probe of New Physics is suggested by
comparison of the direct \CP\ asymmetries in  \BsKpi\ and \BdKpi\ decays.
Using HFAG input~\cite{hfag07} CDF collaboration measured for the first time
  $\frac{\Gamma(\aBdKpi)-\Gamma(\BdKpi)}{\Gamma(\BsKpi)-\Gamma(\aBsKpi)}  
= 0.84 \pm 0.42 \pm 0.15$~\cite{morello_beauty}, in agreement with the Standard Model expectation of unity. 
The branching fraction $\BR(\BsKK)= (24.4 \pm 1.4 \pm 4.6) \times 10^{-6}$ \cite{morello_beauty} 
is in agreement with the latest theoretical expectation~\cite{matiasBsKK,matiasBsKK2}
and with the previous CDF measurement \cite{paper_bhh}.
The results for the \Bd\ are in agreement with world--average values~\cite{hfag07}.
$\acpbdkpi =-0.086 \pm 0.023 \pm 0.009$ is competitive with the current \babelle\ measurements.\\
CDF collaboration also quoted for the first time   
$\BR(\Lbppi)= (3.1 \pm 0.6 \pm 0.7) \times 10^{-6}$ and    
$\BR(\LbpK) = (5.0 \pm 0.7 \pm 1.0) \times 10^{-6}$ \cite{volpi_note}  in agreement with 
theoretical  prediction in \cite{th:BR_lambdab}.   

With full Run II samples (6~\lumifb\ by year 2010) CDF expects  a measurement of \acp\ in \BdKpi\ 
with a statistical plus systematic uncertainty at 1\% level; 5-sigma observation 
of direct \acp\ in \BsKpi\ (or alternatively the possible indication of non-SM sources of CP violation);
more precise measurement of  \acp\ in \Lb\ charmless decays; and 
improved limits, or even observation, of annihilation modes \Bspipi\ and \BdKK.
In addition to the above, time-dependent measurements will be performed for \Bdpipi\ and \BsKK\ decay \cita{CKM06_punzi}.

\section{$\gamma$ from $B\to D K$ decays}

The partial widths of $B^{-} \rightarrow D^0 K^{-}$ modes allow a 
theoretically-clean extraction of the CKM  angle $\gamma = {\rm arg}(-V_{ud}V^*_{ub}/V_{cd}V^*_{cb})$ by a variety of 
methods, depending on the specific $D^0$ decay channel involved 
\cite{glw1,glw2,ads1,ads2,dalitz}.
The precision of current experimental data \cite{hfag07} is still far from theoretical 
uncertainties and is statistics--limited, so the current limited 
knowledge of $\gamma$ can be significantly improved by the addition of further data.
All mentioned methods for extracting $\gamma$ from $B^{-} 
\rightarrow D^0 K^{-}$ modes require no tagging or time-dependent 
measurements, and many of them only involve charged particles in 
the final state. They are therefore particularly well--suited to analysis in a hadronic 
collider environment, where the large production can be well exploited.

CDF collaboration reconstructed modes where the $D^0$ decays in 
either $K^- \pi^+$ (flavor eigenstate) or $K^- K^+, \pi^- \pi^+$
(CP-even eigenstate), in a sample of about 1~\lumifb\ of CDF data.
With a similar analysis of \Bhh\ decays CDF measured the direct CP asymmetry \cite{paola_note}:
$A_{CP+}=0.37\pm 0.14(stat.)\pm 0.04(syst.)$ and the double ratio of the branching fractions:
$R_{CP+} = \frac{R_{+}}{R} = 1.57\pm 0.24(stat.)\pm 0.12(syst.)$ \footnote{For definition 
of $A_{CP+}$ and $R_{CP+}$ see \cite{paola_note}.}
in agreement with previous measurements from other experiments \cite{hfag07}.
This is the first measurement of these quantities to be performed at a hadron collider.

For the search of \BsDsK\ mode CDF collaboration used an unbinned maximum likelihood fit which combines 
the kinematic and  particle identification information.
With a data sample of  integrated luminosity  of $1.2$~\lumifb\ 
CDF observed for the first time \BsDsK\ decays,
with a yield of $109\pm19$ events corresponding to a statistical significance of $7.9\sigma$,
and measured its branching fraction normalized to \BsDspi\ mode
$\frac{\BR(\BsDsK)}{\BR(\BsDspi)}=0.107 \pm 0.019\stat\ \pm 0.007\syst$ \cite{bsdsk_public_note}.
This is the initial step for a possible time-dependent asymmetry measurement
with full Run II statistics. 

\section{New states}

The latest of several observations of new $b$ bound states (after \Lb, \Bc, \Sigmab, \BS, \BSS) is that 
of the \Xib\ baryon in its decay to $\Jpsi\Xi^-$ (with $\Jpsi \to \mu^+\mu^-$ and $\Xi^- \to \Lambda^0 \pi^{-}$)
from the \Do\ and CDF collaborations \cite{Ref:D0_Xib,Ref:CDF_Xib}. With a data sample of 1.3~\lumifb\
\Do\ observed for the first time $15 \pm 4.4\stat\ ^{+1.9}_{-0.4}\syst$ \Xib\ candidates at a mass of 
$5.774 \pm 0.011\stat \pm 0.015~\massgev$ with a significance of 5.5$\sigma$. By normalizing 
to the decay $\Lb \to \Jpsi \Lambda^0$, \Do\ measured the relative rate
$ \frac{\sigma(\Xib)\times \BR(\XibJpsiXi)}{\sigma(\Lb) \times \BR(\Lb \to \Jpsi\Lambda^0) } 
= 0.28 \pm 0.09\stat ^{+0.09}_{-0.08} \syst$.
CDF collaboration, with 1.9~\lumifb\  of data collected,  observed  
$17.5 \pm 4.3$ \Xib\ candidates at a mass of 
$5792.9 \pm 2.5\stat \pm 1.7~\massmev$ with a significance of 7.7$\sigma$.

Heavy quark states provide an interesting laboratory for
testing various approaches to the non-perturbative regime of
QCD.  The Tevatron experiments have made large effort to
improve our knowledge of the $b$-hadrons.

\end{document}